\documentstyle[preprint,aps,epsf]{revtex}
\begin{document}
\input{psfig}
\draft
\title{
\hfill{\normalsize{\bf TRI-PP-97-09, MKPH-T-97-12}}\\[0.3cm]
Generalized polarizabilities and the chiral structure of the nucleon}
\author{Thomas R. Hemmert$^1$, Barry R. Holstein$^2$, Germar
Kn\"{o}chlein$^{1,3}$ and Stefan Scherer$^3$}
\address{$^1$ TRIUMF, Theory Group, 4004 Wesbrook Mall, Vancouver B. C.,
Canada V6T 2A3}
\address{$^2$ Department of Physics and Astronomy,
University of Massachusetts, Amherst MA 01003, USA}
\address{$^3$ Institut f\"{u}r Kernphysik,
Johannes Gutenberg-Universit\"{a}t,
D-55099 Mainz, Germany
}
\maketitle
\begin{abstract}
We present results of
the first chiral perturbation theory calculation
for the generalized polarizabilities of the nucleon and discuss
the response functions in virtual Compton scattering to be measured
in the scheduled electron scattering experiments.
\end{abstract}
\narrowtext
Compton scattering off the nucleon has a long history as a tool for obtaining
precision information on the internal structure of the nucleon.
Beyond the low--energy theorem \cite{LET} the amplitude for
unpolarized Compton scattering
of real photons can be described in terms of two structure--dependent
coefficients \cite{Mac}, commonly denoted as the electric and magnetic
polarizabilities $\bar{\alpha}_E$ and $\bar{\beta}_M$.
   Whereas the constituent quark model traditionally has some problems
\cite{comments} in explaining the numerical values for $\bar{\alpha}_E$
and $\bar{\beta}_M$,
it has been shown that these features can be satisfactorily understood
in chiral perturbation theory (ChPT \cite{ChPTrefs,HBChPT}),
suggesting that $\bar{\alpha}_E$ and $\bar{\beta}_M$ are dominated
by the pion cloud contribution of the nucleon\cite{HBChPT,p4}.

Recently a new frontier in Compton scattering has been defined (see, e.g.,
\cite{VCSprocs}) and is now in the beginning of being explored:
It has been proposed \cite{experiments} to study the electron scattering
process $e p \rightarrow e' p' \gamma$ in order
to obtain information on the genuine virtual Compton
scattering (VCS) process
$\gamma^* N \rightarrow \gamma N$.
In addition to the two kinematical variables
of real Compton scattering, e.g.\ the scattering angle $\theta$ and the
energy $\omega'$ of the outgoing photon,
the invariant amplitude for VCS \cite{Berg,Arenhoevel,Guichon,SKK} depends on
a {\em third} kinematical variable, where we will use the
absolute value of the three--momentum transfer to the nucleon in the
hadronic c.m. frame, $\bar{q}\equiv|\vec{q}|$.
   The structure--dependent coefficients in the VCS amplitude as defined
in \cite{Guichon} acquire a $\bar{q}$ dependence and are termed
``generalized polarizabilities''  (GPs) of the nucleon in analogy to the
structure coefficients in real Compton
scattering. However, due to the specific kinematic approximation chosen
there is no one--to--one correspondence between all
the real Compton polarizabilities and the GPs of Guichon et al.
in VCS \cite{Guichon,formalism,formalism2}.

The virtual nature of the initial state photon with the associated
possibility of
an independent variation of photon energy and momentum allows for
scanning the momentum dependence of the GPs in the electron scattering
experiment, thus rendering access to a much greater variety of structure
information than in the case of real Compton scattering.
   In particular, one hopes to identify the individual signatures of specific
nucleon resonances in the various GPs, which cannot be obtained in other
processes \cite{VCSprocs}.
   In this letter, we will only discuss VCS below the pion--production
threshold.
   For an overview of the status at higher energies and in the
deep inelastic regime we refer to \cite{VCSprocs}.

   So far, predictions for both types of GPs---spin\--independent and
spin\--dependent---have 
been available within a non--relativistic constituent quark model
\cite{Guichon} and a one--loop calculation in the linear sigma model
\cite{Metz}.
Various approaches have been used to calculate
the two spin--independent polarizabilities $\bar{\alpha}_E(\bar{q}^2)$
and $\bar{\beta}_M(\bar{q}^2)$, namely,
an effective Lagrangian approach including nucleon resonance
effects \cite{Vanderhaegen}, 
a calculation of the leading $\bar{q}^2$ dependence 
in heavy--baryon ChPT (HBChPT) \cite{HHKS1} and a 
calculation of $\bar{\alpha}_E(\bar{q}^2)$ in the Skyrme model \cite{KM97}.
   In \cite{formalism,formalism2} it was shown that the number of independent
GPs is reduced, once charge--conjugation symmetry is imposed.

   In this letter we provide the complete set of predictions for the GPs
and the measurable response functions using HBChPT.
   The calculation is performed to third order in the momentum expansion,
which is known to work well in the case of real Compton scattering
\cite{HBChPT}.

The GPs of the nucleon
have been defined in terms of electromagnetic multipoles
as functions of the initial photon momentum $\bar{q}$
\cite{Guichon} ,
\begin{mathletters}
\begin{eqnarray}
\label{gl3_1}
P^{(\rho' L' , \rho L)S} (\bar{q}^2)
& = &
\left[ \frac{1}{\omega'^{L} \bar{q}^{L}}
H^{(\rho' L' , \rho L)S} (\omega' , \bar{q}) \right]_{\omega' = 0} \, ,
\\ \label{gl3_1b}
\hat{P}^{(\rho' L' , L)S} (\bar{q}^2) & = &
\left[ \frac{1}{\omega'^{L} \bar{q}^{L+1}}
\hat{H}^{(\rho' L' , L)S} (\omega' , \bar{q}) \right]_{\omega' = 0}
\, ,
\end{eqnarray}
\end{mathletters}
\noindent where
$L$ ($L'$) denotes the initial (final) photon orbital angular momentum,
$\rho$ ($\rho'$) the type of multipole transition ($0 = C$ (scalar, Coulomb),
$1 = M$ (magnetic),
$2 = E$ (electric)), and $S$ distinguishes between non--spin--flip ($S=0$) and
spin--flip ($S=1$) transitions.
Mixed--type polarizabilities,
${\hat{P}}^{(\rho' L' , L)S} (\bar{q}^2)$, have been introduced,
which are neither purely electric nor purely Coulomb type.
It is important to note that the above definitions are based on the
kinematical approximation
that the multipoles are expanded around $\omega' = 0$
and {\em{only terms linear
in $\omega'$ are retained}}, which together with current conservation yields
selection rules for the possible combinations of quantum numbers of the 
GPs. In this approximation, 10 GPs
have been introduced in \cite{Guichon}
as functions of $\bar{q}^2$:
$P^{(01,01)0}$,
$P^{(11,11)0}\,$,
$P^{(01,01)1}\,$,
$P^{(11,11)1}\,$,
$P^{(01,12)1}\,$,
$P^{(11,02)1}\,$,
$P^{(11,00)1}\,$,
${\hat{P}}^{(01,1)0}\,$,
${\hat{P}}^{(01,1)1}\,$,
${\hat{P}}^{(11,2)1}\,$.

However, recently it has been proved \cite{formalism,formalism2} that only
six of the above ten GPs are independent,
if one requires charge--conjugation symmetry to hold.
   With $\omega_0 \equiv M - \sqrt{M^2+\bar{q}^2}$, $M$ being the nucleon mass,
the four constraints implied by $C$ invariance can be written as
\begin{mathletters}
\label{gl3_6}
\begin{eqnarray} \label{gl3_6a}
0 & = & \sqrt{\frac{3}{2}} P^{(01,01)0}(\bar{q}^2)
 + \sqrt{\frac{3}{8}} P^{(11,11)0}(\bar{q}^2)
 + \frac{3 \bar{q}^2}{2 \omega_0} \hat{P}^{(01,1)0}(\bar{q}^2) \, ,
\\ \label{gl3_6b}
0 & = & P^{(11,11)1}(\bar{q}^2)
 + \sqrt{\frac{3}{2}} \omega_0 P^{(11,02)1}(\bar{q}^2)
 + \sqrt{\frac{5}{2}} \bar{q}^2 \hat{P}^{(11,2)1}(\bar{q}^2) \, ,
\\ \label{gl3_6c}
0 & = & 2 \omega_0 P^{(01,01)1}(\bar{q}^2)
 + 2 \frac{\bar{q}^2}{\omega_0} P^{(11,11)1}(\bar{q}^2)
 - \sqrt{2} \bar{q}^2 P^{(01,12)1}(\bar{q}^2)
 + \sqrt{6} \bar{q}^2 \hat{P}^{(01,1)1}(\bar{q}^2) \, ,
\\ \label{gl3_6d}
0 & = & 3 \frac{\bar{q}^2}{\omega_0} P^{(01,01)1}(\bar{q}^2)
 - \sqrt{3} P^{(11,00)1}(\bar{q}^2)
 - \sqrt{\frac{3}{2}} \bar{q}^2 P^{(11,02)1}(\bar{q}^2) \, .
\end{eqnarray}
\end{mathletters}

In the scalar (i.e. spin--independent) sector
it is convenient to use Eq.(\ref{gl3_6a}) to eliminate the mixed
polarizability ${\hat{P}}^{(01,1)0}$
in favor of $P^{(01,01)0}$ and $P^{(11,11)1}$, because the latter are
generalizations of the electric and magnetic polarizabilities in real
Compton scattering:
\begin{mathletters}
\begin{eqnarray} \label{gl3_2a}
\bar{\alpha}_E (\bar{q}^2) & = & - \frac{e^{2}}{4 \pi} \sqrt{\frac{3}{2}}
P^{(01,01)0} (\bar{q}^2) \,,
\\ \label{gl3_2b}
\bar{\beta}_M (\bar{q}^2) & = & - \frac{e^{2}}{4 \pi} \sqrt{\frac{3}{8}}
P^{(11,11)0} (\bar{q}^2) \,.
\end{eqnarray}
\end{mathletters}
However, in the spin--dependent sector it is not a priori clear which
three GPs should be eliminated with the help of the
constraints---Eqs.\ (\ref{gl3_6b})-(\ref{gl3_6d}). Thus, and for easier
comparison with
other theoretical calculations which have been performed before the
constraints from Eq.\ (\ref{gl3_6}) were recognized, we will
give results
for the original set of 10 GPs.
   Eqs.\ (\ref{gl3_6a}) - (\ref{gl3_6d}) then provide a useful check for
any calculation of the GPs.

   Our calculation of the GPs is performed within
HBChPT \cite{HBChPT} to third order in the external momenta --- $O(p^3)$.
The chiral results are highly constrained, the
only parameters being
the pion mass $m_\pi$,
the axial coupling constant $g_A$, and the pion
decay constant $F_\pi$, which are all known very accurately.
   At $O(p^3)$, contributions to the GPs are
generated by nine one--loop diagrams and the $\pi^0$--exchange
$t$--channel pole graph which are displayed in \cite{HHKS1}.
   For the loop diagrams only the leading--order Lagrangians are
required \cite{HBChPT}:
\begin{mathletters}
\label{lagrangian}
\begin{eqnarray}
{\cal{L}}_{\pi N}^{(1)} & = & \bar{N}_v (i v \cdot D + g_A S \cdot u) N_v
\,, \\
{\cal{L}}_{\pi \pi}^{(2)} & = &
\frac{F_{\pi}^2}{4} {\mathrm{tr}} \left[ (\nabla_{\mu} U)^{\dagger}
\nabla^{\mu} U \right]
\,.
\end{eqnarray}
\end{mathletters}
   In the above $SU(2)$ Lagrangians, $N_v$ represents a non-relativistic
nucleon field, and $U = {\mathrm{exp}}(i \vec \tau \cdot \vec \pi/F_{\pi})$
contains the pion field.
   Standard covariant derivatives have been introduced acting on the
pion ($\nabla_{\mu}U$) and on the nucleon ($D_{\mu} N_v$), in addition to the
chiral vielbein $u_\mu$, which contains the derivative coupling of a pion.
   In the heavy--baryon Lagrangian one also must
specify the velocity vector $v_\mu$, which determines the representation of
the Pauli--Lubanski spin--vector
$S^{\mu} = i \gamma_5 \sigma^{\mu\nu} v_{\nu}$ \cite{HBChPT}.
   Finally, for the $\pi^0$--exchange diagram we require in addition to
Eq.\ (\ref{lagrangian}) the anomalous $\pi^0\gamma\gamma^\ast$ vertex
provided by the Wess-Zumino-Witten Lagrangian \cite{WZW},
\begin{equation}
\label{wzwpi0}
{\cal{L}}_{\gamma\gamma\pi^0}^{(WZW)} =  -\frac{e^2}{32\pi^2 F_\pi} \;
\epsilon^{\mu\nu\alpha\beta} F_{\mu\nu} F_{\alpha\beta} \pi^0 \,,
\end{equation}
   where $\epsilon_{0123}=1$ and $F_{\mu\nu}$ is the electromagnetic field
strength tensor.
   Details of the calculation will be reported in a separate publication.

   In \cite{HHKS1} we have already performed a calculation of the first
derivative of the spin--independent polarizabilities with respect to
${\bar{q}}^2$, and technically the calculation of the material presented here 
proceeds along the same line.
However, \cite{HHKS1} focuses on the VCS regime of
very low momentum transfer ($\omega',\bar{q} \ll m_\pi$),
for which the formalism introduced in \cite{Guichon} is not applicable.
   The extraction of the GPs from a heavy--baryon calculation is described in
detail in Sec. IV of \cite{formalism2}.
The numerical results for the ten generalized proton
polarizabilities are presented in Fig.\ref{gpols}.
Therein, the contribution from the anomalous Wess--Zumino--Witten interaction
only arises in the spin--dependent sector and is plotted separately.

   Let us first discuss the electric polarizability
$\bar{\alpha}_E(\bar{q}^2)$, which is numerically by far the largest of the
ten GPs and thus can presumably be determined most
reliably in the scheduled experiments.
At the kinematic point $\bar{q}^2=0$ the generalized electric polarizability
$\bar{\alpha}_E(\bar{q}^2)$ can be identified
with the electric polarizability $\bar{\alpha}_E$ from real Compton scattering
as discussed in \cite{Guichon,formalism}.
   The prediction of the $O(p^3)$ calculation for
$\bar{\alpha}_E(0)=\bar{\alpha}_E$ agrees remarkably well with the
experimental value extracted from real
Compton scattering (HBChPT [O($p^3$)]: $12.8 \times 10^{-4}\mbox{fm}^3$;
Expt.: $(12.1 \pm 0.8 \pm 0.5)\times 10^{-4}\mbox{fm}^3$
\cite{Mac}).
   Slowly increasing three--momentum transfer $\bar{q}$, ChPT predicts
a sharp decrease for the generalized electric polarizability
$\alpha_E (\bar{q}^2)$ \cite{HHKS1}, which is
considerably faster than predicted by the constituent quark model
\cite{Guichon}.

Likewise, we present the full $\bar{q}^2$ evolution of the generalized
magnetic polarizability $\bar{\beta}_M(\bar{q}^2)$ in Fig.\ref{gpols}.
At $\bar{q}^2=0$, it is falling short
of the central experimental value from real Compton scattering
(HBChPT [O($p^3$)]: $1.3 \times 10^{-4}\mbox{fm}^3$; Expt.: $(2.1 \mp 0.8
\mp 0.5) \times 10^{-4}\mbox{fm}^3$ \cite{Mac}). It is known that this 
discrepancy can be resolved in a $O(p^4)$ ChPT calculation, as at that order 
one can implement the effects of the nucleon resonance $\Delta$(1232) 
\cite{p4}. In our $O(p^3)$ calculation, ChPT predicts 
$\bar{\beta}_M(\bar{q}^2)$ to rise at small momentum transfer, 
with a {\it turnover point} around $\bar{q}^2=0.1\mbox{GeV}^2$. Possible
higher order modifications of this $\bar{q}^2$ behavior due to 
$\Delta$(1232) are currently under investigation \cite{HHK}. 

Finally, let us address the seven spin--dependent GPs, constrained by 
Eqs.(\ref{gl3_6b})-(\ref{gl3_6d}). At this point, there 
exists not even a measurement of the corresponding polarizabilities in real
Compton scattering to compare with. The possibility
of such measurements is being studied \cite{rory}. We also note that the
familiar ``spin-polarizability'' $\gamma$ of the nucleon \cite{HBChPT,p4} is
not accessible in the kinematical approximation of \cite{Guichon} as discussed
in \cite{formalism2}.
Our predictions for the spin--dependent GPs, shown in 
Fig.\ref{gpols}, contain two rather distinct
contributions---the first originating from pionic loop contributions,
whereas the second is associated with the anomalous
$\pi^0 \gamma\gamma^\ast$ vertex. 
   It is interesting that at $O(p^3)$ the contributions of the 
pion--nucleon loops to the spin--dependent GPs are much smaller
than the contributions arising from the $\pi^0$--exchange diagram !

\begin{figure}
\caption{
\small
Response functions in $e p \rightarrow e' p' \gamma$. The dashed line
results from a dipole--parameterization of the
proton form factors with $G_D(Q^2)=[1+Q^2/(0.71\mbox{GeV}^2)]^{-2}$,
whereas the dotted curve utilizes the Taylor expanded formfactors 
$G_T(Q^2)=1 + Q^2 \; \frac{dG_D}{d Q^2}|_{Q^2=0}$. 
The difference between the two curves represents a 
measure for the importance of $O(p^n), \; n>3$ corrections.
\protect\label{resp}
}
\end{figure}
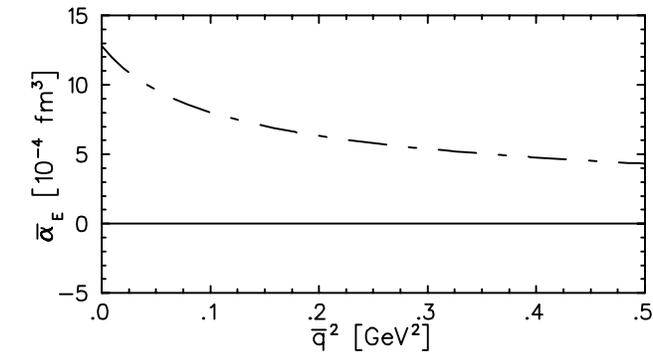
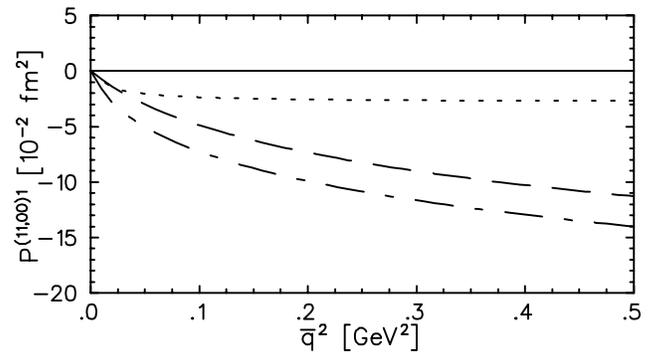
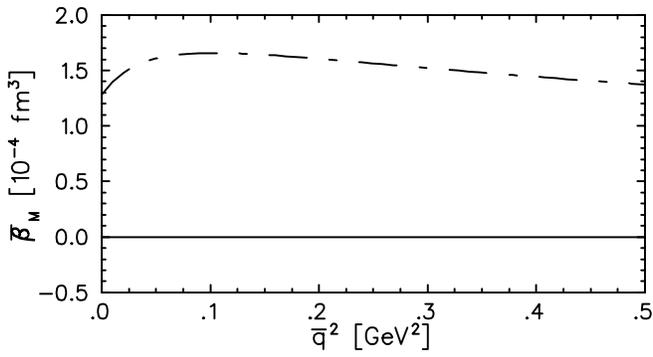
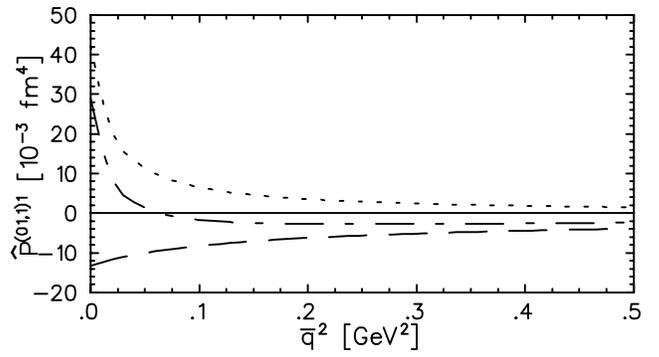
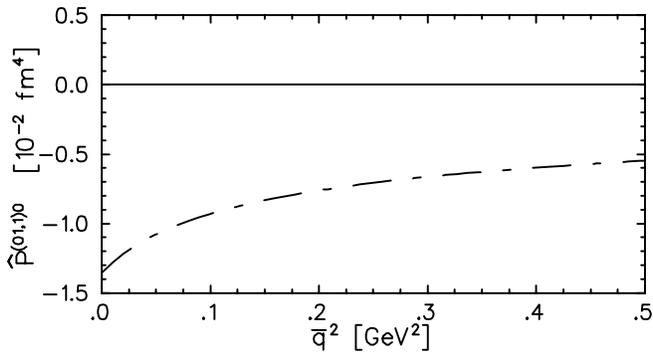
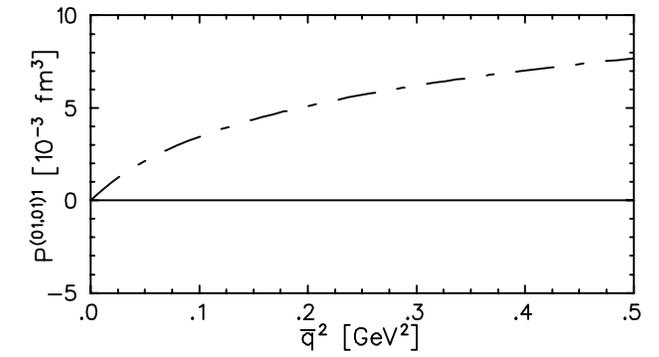
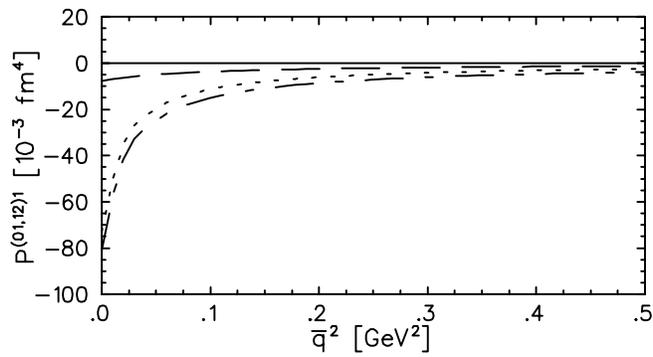
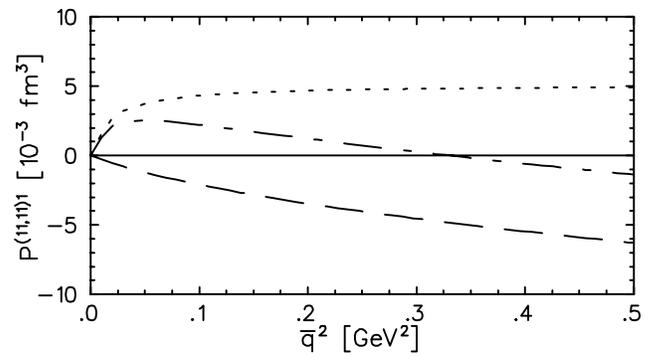
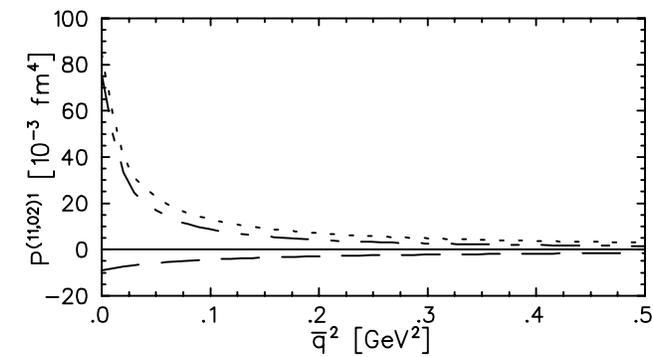
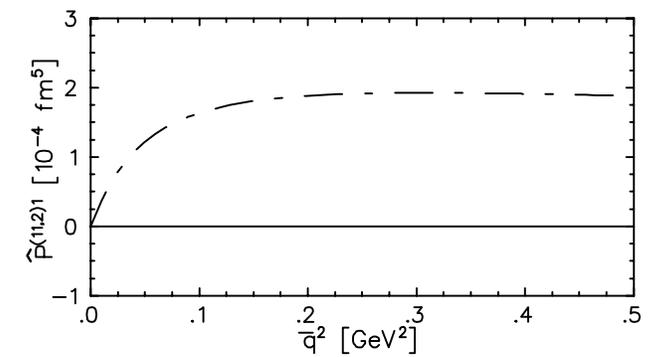
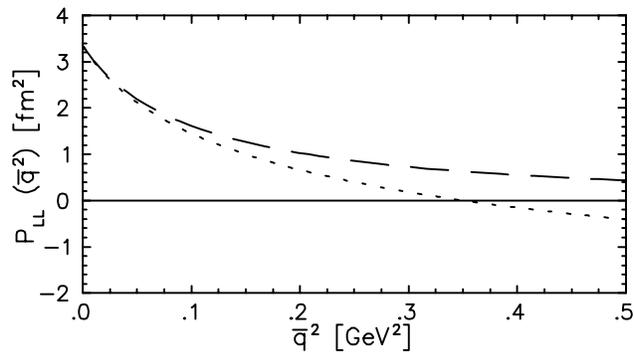
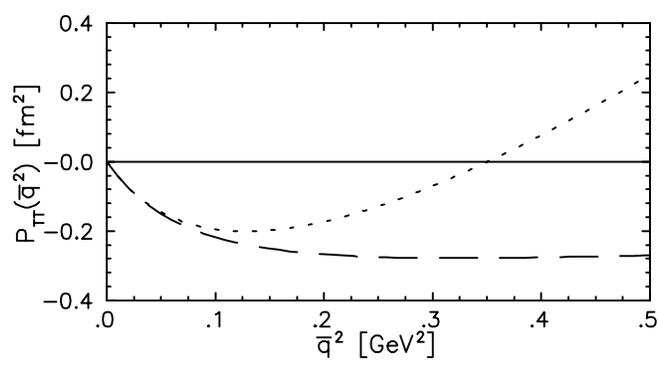
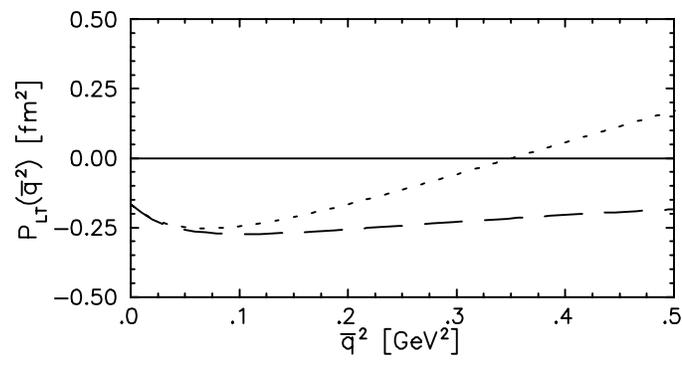

Experimentally, the extraction of the generalized (proton) polarizabilities
is accomplished by utilizing the interference of the
structure--dependent part of the VCS matrix element with
the Bethe--Heitler and nucleon pole diagrams.
Once $C$ invariance is imposed, in the kinematical approximation of Guichon
et al. \cite{Guichon} the lowest--order structure--dependent
contributions in an {\em unpolarized experiment} can then be
parameterized in terms of {\em three independent response functions}
\cite{formalism2},
each containing products of nucleon form factors with the GPs
as building blocks:

\begin{mathletters}
\begin{eqnarray} \label{gl3_10a}
P_{LL}(\bar{q}^2) & = &
-2 \sqrt{6} M G_{E}(Q_{0}^{2}) P^{(01,01)0}(\bar{q}^2) \, ,
\vphantom{\frac{1}{1}}
\\ \label{gl3_10b}
P_{TT}(\bar{q}^2) & = &
\frac{3}{2} G_{M}(Q_{0}^{2})
\Bigl [ 2 \omega_{0} P^{(01,01)1}(\bar{q}^2)
+ \sqrt{2} \bar{q}^{2} \Bigl ( P^{(01,12)1}(\bar{q}^2)
              + \sqrt{3} \hat{P}^{(01,1)1}(\bar{q}^2) \Bigr ) \Bigr ] \, ,
\\ \label{gl3_10c}
P_{LT}(\bar{q}^2) & = &
\sqrt{\frac{3}{2}} \frac{M \bar{q}}{\sqrt{Q_{0}^{2}}}
G_{E}(Q_{0}^{2}) P^{(11,11)0}(\bar{q}^2)
+ \frac{\sqrt{3} \sqrt{Q_{0}^{2}}}{2 \bar{q}} G_{M}(Q_{0}^{2})
\Bigl ( P^{(11,00)1}(\bar{q}^2)
+ \frac{\bar{q}^{2}}{\sqrt{2}} P^{(11,02)1}(\bar{q}^2) \Bigr ) \, .
\label{gl3_10d}
\end{eqnarray}
\end{mathletters}
Here $G_E (Q^2)$ ($G_M(Q^2)$) are the electric (magnetic) nucleon
Sachs form factors and $Q_0^2 = Q^2|_{\omega'=0}$. The ChPT
predictions for the response functions are given in Fig.\ref{resp}. 
We note that
the particular kinematic approximation suggested in \cite{Guichon}
has the remarkable property that the $\pi^0$ pole
contributions cancel exactly {\it at
the level of the response functions} in an $O(p^3)$ calculation.
Therefore, the information displayed in Fig.\ref{resp} is directly based
upon the pion--nucleon loop effects and the parameterization of the nucleon
form factors. We have chosen to use a Taylor expansion of the nucleon
form factors up to the first $Q^2$ coefficient in addition to the
standard dipole--form. In
Fig.\ref{gpols} one can see the differences between
the parameterizations, which we take as a measure of the uncertainty of
the $O(p^3)$ ChPT prediction. According to our analysis, the response
function $P_{LL}(\bar{q}^2)$ is the one that can be predicted most reliably.
It directly allows for extracting $\bar{\alpha}_E(\bar{q}^2)$.

The Mainz experiment (G. Audit et al. \cite{experiments}) will supply data at
$\bar{q}=600 MeV$, which
can be used to determine $\bar{\alpha}_E({\bar{q}}^2 = 0.36 GeV^2)$ and two
additional linear combinations of the GPs. The $O(p^3)$ ChPT calculation
may not be valid at this ``large'' momentum transfer, but Fig.\ref{resp}
predicts that $\bar{\alpha}_E$ should have
decreased by $\approx 50\%$ from its real--photon value. It will be
interesting to collect data at smaller ${\bar{q}}^2$, as it is planned in
the $\bar{\alpha}_E$ optimized experiment at
MIT-Bates ( J. Shaw et al. \cite{experiments}). At this relatively small
momentum transfer ($\bar{q} \approx 240$MeV) the $O(p^3)$ HBChPT calculation
of the GPs should be most reliable. However, when
$\bar{q}$ approaches the order of magnitude of $\omega'$ the response function
analysis of \cite{Guichon} will break down, due to the
higher--order terms in
$\omega'$ which are neglected therein. For even smaller $\bar{q}$
one then approaches the kinematical regime that has been dealt with in
\cite{HHKS1}. In summary the two above experiments taken together with the
value of $\alpha_E(0)$
provide the first opportunity to determine the $\bar{q}^2$ dependence of
a GP.

We have presented the first complete
ChPT calculation for the GPs of the nucleon,
which are going to be measured in several upcoming experiments.
Future work will include an $O(p^4)$ calculation and the consideration of 
explicit $\Delta(1232)$ degrees of freedom.
Work in these directions is in progress \cite{HHK}.

Research supported in part by NSERC, NSF, DAAD(HSPIII) and DFG(SFB201).

\newpage
\begin{figure}
\caption{
\small
$O(p^3)$ prediction for the GPs of the proton as a function of $\bar{q}^2$. 
The dashed line represents the
contribution from pion--nucleon loops, the dotted one from the $\pi^0$
exchange graph and the dash--dotted line the sum of both.
\protect\label{gpols}
}
\end{figure}

\end{document}